\documentclass[9pt]{osajournal}
\setboolean{shortarticle}{false}
\usepackage{amsmath,bm}
\usepackage{xspace}
\usepackage{multicol}

\def\resultPath{Figures/results/}
\newcommand{\mymath}[2]{\newcommand{#1}{\TextOrMath{$#2$\xspace}{#2}}}
\newcommand{\eg}{e.\,g.,\ }
\newcommand{\ie}{i.\,e.,\ }
\newcommand{\object}{\mathrm{O}}
\newcommand{\detector}{\mathrm{D}}
\mymath{\generator}{\bm G}
\mymath{\discriminator}{\bm D}
\mymath{\phaseDiscriminator}{{\discriminator_\object}}
\mymath{\intensityDiscriminator}{{\discriminator_\detector}}
\mymath{\propagator}{\bm H}
\mymath{\objectGenerator}{\generator_\object}
\mymath{\detectorGenerator}{\generator_\detector}
\mymath{\objectWaveField}{\bm\psi_\object}
\mymath{\detectorWaveField}{\bm\psi_\detector}
\mymath{\propagatedIntensity}{|\bm\psi_\detector|^2}
\mymath{\detectorIntensity}{\bm I_\detector}
\mymath{\ganLoss}{\mathcal L_\mathrm{GAN}}
\mymath{\cycleLoss}{\mathcal L_\mathrm{Cyc}}
\mymath{\frcLoss}{\mathcal L_\mathrm{FRC}}
\mymath{\ganLossWeight}{\alpha_\mathrm{GAN}}
\mymath{\cycleLossWeight}{\alpha_\mathrm{Cyc}}
\mymath{\frcLossWeight}{\alpha_\mathrm{FRC}}
\mymath{\mseLossWeight}{\alpha_\mathrm{MSE}}
\mymath{\phaseSamples}{\bm \Psi}
\mymath{\intensitySamples}{\mathcal I}
\mymath{\frc}{\operatorname{\bf{ FRC}}}
\newcommand{\method}[1]{\texttt{#1}}
\newcommand{\icon}[1]{\raisebox{-.5\height}{\includegraphics[width = 0.09\textwidth]{\resultPath #1}}}
\newcolumntype{P}[1]{>{\centering\arraybackslash}p{#1}}
\newcommand{\methodInTable}[1]{%
\rotatebox[origin=c]{90}{\method{#1}}&%
\icon{#1_crop_re} &%
\icon{#1_crop_im} &%
}
\newcommand{\minihisto}[2]{%
\multirow{1}{1.2cm}{%
    $\mu=#1$\\%
    \includegraphics[width = 0.06\textwidth]{Figures/histograms/#2}%
}%
}

\usepackage[nolist]{acronym}
\usepackage{comment}
\usepackage{multirow}
\usepackage{makecell}
\usepackage{graphicx}
\usepackage[export]{adjustbox}

\begin{acronym}
\acro{APS}{Advanced Photon Source}
\acro{CTF}{Contrast Transfer Function}
\acro{DL}{Deep Learning}
\acro{FRC}{Fourier Ring Correlation}
\acro{FRCM}{Fourier Ring Correlation Metric}
\acro{GAN}{Generative Adversarial Network}
\acro{NN}{Neural Network}
\acro{PCA}{Principal Component Analysis}
\acro{DSSIM}{Dissimilarity Structure Similarity Index Metric}
\acro{TIE}{transport of intensity equations}
\acro{XFEL}{X-ray free-electron laser}
\acro{GPU}{Graphical Processing Unit}
\end{acronym}

\setboolean{shortarticle}{false}
\title{PhaseGAN: A deep-learning phase-retrieval approach for unpaired datasets}

\author[1,*]{Yuhe Zhang}
\author[2]{Mike Andreas Noack}
\author[3]{Patrik Vagovic}
\author[4]{Kamel Fezzaa}
\author[5]{Francisco Garcia-Moreno}
\author[6]{Tobias Ritschel}
\author[1]{Pablo Villanueva-Perez}

\affil[1]{Synchrotron Radiation Research and NanoLund, Lund University, Box 118, 221 00, Lund, Sweden}
\affil[2]{Technische Universität Berlin, 10623 Berlin, Germany}
\affil[3]{Center for Free-Electron Laser Science, DESY, 22607 Hamburg, Germany}
\affil[4]{X-ray Science Division,Advanced Photon Source,Argonne National Laboratory, Lemont, IL 60439,US}
\affil[5]{Helmholtz-Zentrum Berlin für Materialien und Energie, 14109 Berlin, Germany}
\affil[6]{University College London, WC1E 6BT London, UK}
\affil[*]{Corresponding author: yuhe.zhang@sljus.lu.se}

\begin{abstract}
Phase retrieval approaches based on \ac{DL} provide a framework to obtain phase information from an intensity hologram or diffraction pattern in a robust manner and in real time. However, current \ac{DL} architectures applied to the phase problem rely i) on paired datasets, \ie they are only applicable when a satisfactory solution of the phase problem has been found, and ii) on the fact that most of them ignore the physics of the imaging process. Here, we present PhaseGAN, a new \ac{DL} approach based on Generative Adversarial Networks, which allows the use of unpaired datasets and includes the physics of image formation. Performance of our approach is enhanced by including the image formation physics and provides phase reconstructions when conventional phase retrieval algorithms fail, such as ultra-fast experiments. Thus, PhaseGAN offers the opportunity to address the phase problem when no phase reconstructions are available, but good simulations of the object or data from other experiments are available, enabling us to obtain results not possible before.
\end{abstract}

\setboolean{displaycopyright}{true}

\begin{document}
\maketitle

\section{Introduction}

Phase retrieval, \ie reconstructing phase information from intensity measurements, is a common problem in coherent imaging techniques such as holography~\cite{Gabor1948}, coherent diffraction imaging~\cite{Miao1999}, and ptychography~\cite{Hegerl1970,Faulkner2004}.
As most detectors only record intensity information, the phase information is lost, making its reconstruction an ill-defined problem~\cite{Fienup1982,TEAGUE1982}.
Most common quantitative solutions to the phase problem either rely on deterministic approaches or on an iterative solution~\cite{ZUO2020106187}.
Examples of deterministic solutions to holography are \ac{TIE}~\cite{ReedTeague1983a} or based on  \acp{CTF}~\cite{Guigay1977}.
Such deterministic approaches can only be applied if certain constraints are met.
For example, \ac{TIE} is valid only in paraxial and short-propagation-distance conditions.
Furthermore, complex objects can only be reconstructed with \ac{TIE} when assuming a spatially homogeneous material~\cite{Paganin2002a}.
Similarly, \ac{CTF} only applies to weak scattering and absorption objects.
Iterative approaches are not limited by these constraints~\cite{Gerchberg1972a,Fienup1978} and can address not only holography but also coherent diffraction imaging and ptychography.
These techniques retrieve the object by alternating between the detector and object space and iteratively applying constraints on both spaces, as depicted in Fig.~\ref{fig:Schema}(a).
This process is computationally expensive, requiring several minutes to converge, precluding application to real-time analysis.
Furthermore, the convergence of such approach is not guaranteed.

Recently, \ac{DL} has demonstrated potential to solve ill-posed imaging problems, such as holography~\cite{Rivenson2018,Wang2018}, magnetic resonance imaging~\cite{Leiner2019}, and phase retrieval~\cite{Sinha2017,Cherukara2018}.
\ac{DL} offers an accurate solution to the phase problem, which is computationally fast compared to iterative approaches~\cite{Rivenson2018,Cherukara2018}, and independent of physical approximations.
\ac{DL} methods need to be \emph{trained} before they are used.
In \emph{supervised} training, data is input to a differentiable method with adjustable parameters, \eg a \ac{NN}.
The \ac{NN} is then \emph{tuned} to produce the desired output.
In classic \emph{paired} supervision applied to phase retrieval, for every input (intensity) such training needs to know the precise output (phase).
This paired supervision has two main difficulties:

First, such approaches require recording large datasets of phase and intensity of exactly the same sample.
It is easy to think of conditions where this is not possible:
i) Some instruments, like \ac{XFEL}~\cite{Madey1971FEL,Kondratenko1980XFEL,Bonifacio85XFEL,Bonifacio1994XFEL} have limited accessibility, making it difficult to acquire large paired datasets from such instruments.
ii) Phase retrieval algorithms might not provide good reconstructions or are not even applicable.
Examples of such scenarios are diffraction experiments where only simulations are available but not phase reconstructions~\cite{Davtyan2017DiffRecoSim,Diaz2010BraggSimReco} or Bragg Coherent Diffraction imaging~\cite{Robinson2009BraggCDI} experiments where obtaining good phase reconstructions have proven a challenging task~\cite{Carnis2019ProblemBCDI,Wang2020BraggCDIProbl}.
iii) Complementary imaging modalities, \eg certain imaging experiments might provide low-noise and high-spatial-resolution phase reconstructions while another experiment provides high-noise detector images at a lower resolution of similar samples, but not of the same exact sample.
This is of particular importance when imaging radio-sensitive samples with directly or indirectly-ionizing radiation, such X-rays. Such scenario requires minimizing the deposited dose, \ie deposited energy per unit of mass.
Alternatively, this is a typical problem when performing fast imaging experiments to track dynamics with a reduced number of photons per exposure.
iv) Sensing might alter or even destroy the sample, \eg in a diffraction-before-destruction imaging modality with high-intensity sources such as \acp{XFEL}~\cite{Neutze2000DiffBDest,Chapman2006DiffBDest}. In this scenario, rendering paired sensing with a different modality is impossible.
We argue how \emph{unpaired} training, where all we need is random samples from the two different experimental setups, but not from the same object, will overcome all these four (i--iv) limitations.

Second, even if paired data was available, the results are often unsatisfying when attempting to solve an ill-posed problem, \ie if one intensity reading does not map to one specific phase solution~\cite{Bruck1979} but to a distribution of possible explanations.
Classic paired training is known to \emph{average}, \ie spatially blur, all possible solutions if the solution is not unique \cite{ledig2017photo}.
Adversarial training~\cite{goodfellow2014generative} can overcome this problem by augmenting the training by a \emph{discriminator}, \ie another \ac{NN}, with the purpose to correctly classify results from the training, as well as true samples of the data distribution, \ie from-the-wild phase images, as either ``real'' or ``fake''.
The training uses the information of what was objectionable so that the discriminator could detect a method's results as fake, to improve the method itself.
It also uses the information from the true samples of the data distribution to become picky, \ie good at saying what is ``real'' or ``fake''.
For ill-posed problems such as phase reconstruction, this will push the solution away from the average of all possible phase images that explain an intensity image ---which itself would not be a good phase image, as it is blurry--- to a specific solution, which also explains the input, but is not blurry.

New \ac{DL} adversarial schemes have shown the possibility of training on unpaired data sets; that is, a set of images captured from one modality and another set made using a different modality, but not necessarily of the same object.
CycleGAN~\cite{zhu2017unpaired} learns a pair of \emph{cycle consistent} functions, which map from one modality to the other such that their composition is the identity.
This consistency constraint is analogous to the constraint applied in iterative phase reconstruction algorithms~\cite{Gerchberg1972a,Fienup1982}, where cyclic constraints are applied between the sample and detector space.
Thus, approaches based on CycleGAN offer a framework for phase reconstruction, which mimics the structure of iterative approaches but without the limitation to paired datasets.

In this paper, we demonstrate a \ac{DL} implementation, christened PhaseGAN, based on CycleGAN.
PhaseGAN naturally includes the physics of the image formation as it cycles between the sample and the detector domains.
By including the physics of the image formation and other learning constraints, PhaseGAN retrieves phase reconstructions better than CycleGAN, which are comparable to state-of-the-art paired approaches.

The remainder of this paper is structured as follows:
First, we describe our approach's architecture and how the physics of the image formation is included.
Second, we validate PhaseGAN with synthetic data for in-line holographic (near-field) experiments. In this validation step, we demonstrate the relevance of including the physical model by comparing the results with CycleGAN. Furthermore, we demonstrate that our unpaired approach performs at the level of state-of-the-art paired approaches.
Third, we apply PhaseGAN to fast-imaging experimental data where noisy readings of a MHz camera are reconstructed using low-noise phase reconstructions recorded with a different setup and samples.
Finally, we discuss the results and future applications of PhaseGAN to experiments where phase reconstructions are not possible today.

\section{The PhaseGAN approach}
\label{sec:method}

\begin{figure}[htb]
\centering\includegraphics*[width = 0.5\linewidth]{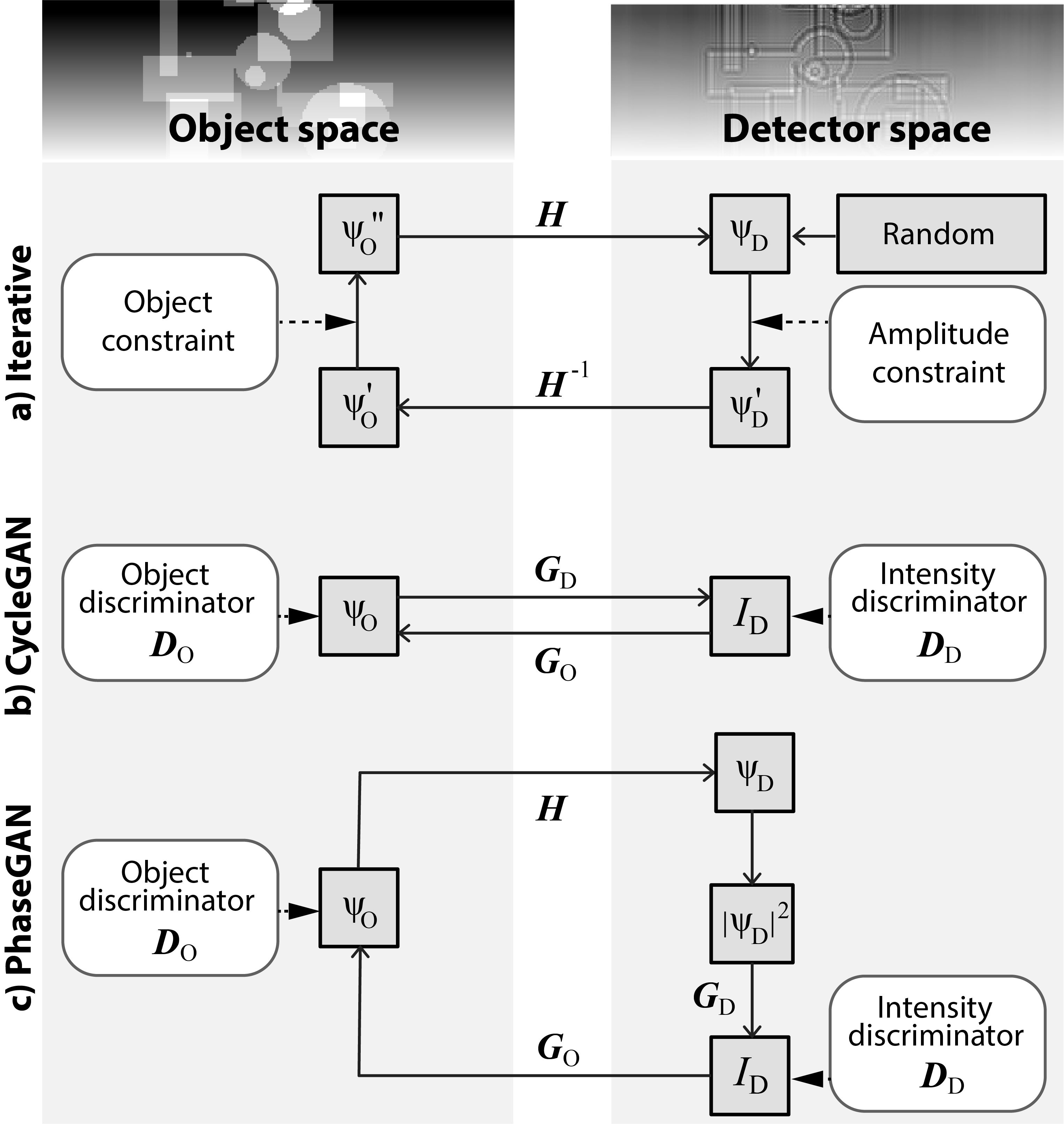}
\vspace{-.2cm}
\caption{Schematic approach of  (a) conventional iterative phase-retrieval approaches, (b) CycleGAN, and (c) PhaseGAN.}
\label{fig:Schema}
\end{figure}

This section describes the architecture of PhaseGAN and how it uses physical knowledge to enhance the phase reconstructions.
We then describe the training process and our loss function, which includes terms that avoid typical phase-reconstruction artifacts such as missing frequencies or the twin-imaging problem~\cite{Gabor1948,Latychevskaia2007TwinImage}.

\begin{figure*}[htb]
\centering\includegraphics*[width = \linewidth]{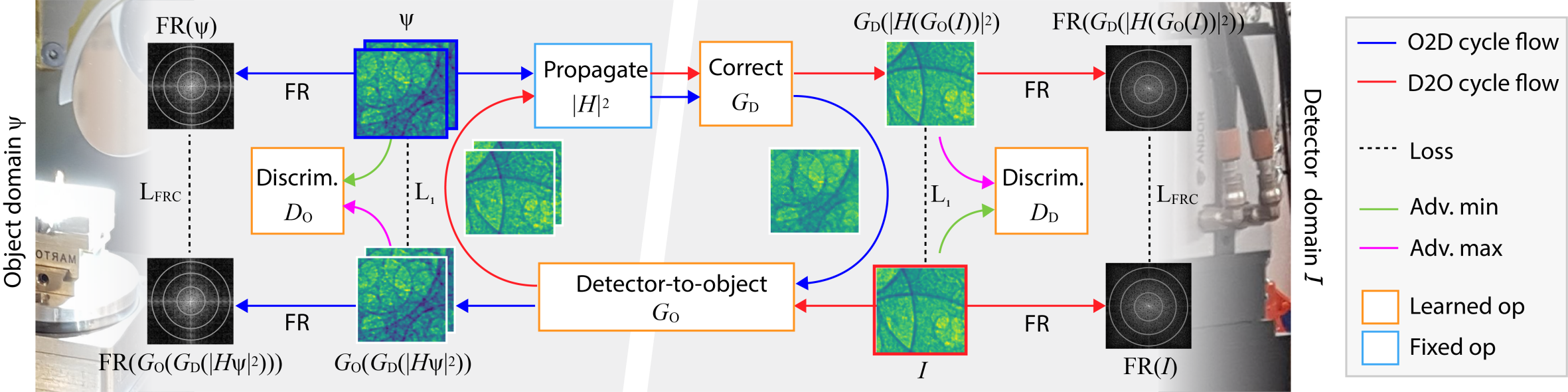}
\vspace{-.2cm}
\caption{Learning process diagram. Our aim is to learn a mapping $\objectGenerator$ from the intensity sensing regime \textbf{(right)} to a phase modality \textbf{(left)}.
We require this mapping $\objectGenerator$ to fulfill two cyclic constraints:
First \textbf{(blue)}, when its phase result is being mapped back to the intensity domain using a non-learned physical operator $\propagator$ and a learned correction operation $\detectorGenerator$, the result should be similar \textbf{(dotted line)} to the intensity.
Second \textbf{(red)}, when the phase is mapped to intensity and back, it should remain the same. Further, we train two discriminators $\intensityDiscriminator$ and $\phaseDiscriminator$ to classify real and generated intensity and phase samples as real or fake \textbf{(green)}.
Finally, we ask the Fourier transform, another fixed but differentiable op of both intensity and phase, to match the input after one cycle. }
\label{fig:Main}
\end{figure*}

The architecture of PhaseGAN is based on CycleGAN~\cite{zhu2017unpaired}.
CycleGAN uses two \ac{GAN}s, which allow the translation of one image from a domain $A$ to a domain $B$ and the inverse translation from $B$ to $A$.
Thus, the cycle consistency between two domains can be adapted to the object and detector domains, allowing CycleGAN to perform phase reconstructions by mimicking the structure of iterative phase-retrieval approaches, as shown in Fig.~\ref{fig:Schema}(b).
The main difference between iterative phase-retrieval approaches and CycleGAN approaches is the inclusion of the propagator (\propagator), which includes the physics of the image formation between the object and the detector space.
PhaseGAN combines both the iterative and the CycleGAN approach by including two \ac{GAN}s in a cyclic way together with the physics of the image formation via the propagator.
The scheme of PhaseGAN is depicted in Fig.~\ref{fig:Schema}(c), where each of the \acp{GAN} is decomposed in their generator (\generator) and discriminator (\discriminator).
The generators used in PhaseGAN are U-Net~\cite{ronneberger2015u}-like end-to-end fully convolutional neural networks.
For specific details about the generators see \href{link}{Supplement 1}.
The discriminators are PatchGAN discriminators ~\cite{ledig2017photo,zhu2017unpaired}.
\objectGenerator~is the phase-reconstruction generator, which takes the measured intensities (one single channel input) and produces a two-channel complex output, where the two channels can be either the real and imaging part or the phase and amplitude of the complex-object wave field (\objectWaveField). 
\phaseDiscriminator is the discriminator of the phase reconstruction.
The object wavefield \objectWaveField is then propagated using the non-learnable but differentiable operator \propagator to the detector plane ($\detectorWaveField=\propagator\objectWaveField$), and the intensity in the detector plane is computed (\propagatedIntensity).
The propagator \propagator is the near-field Fresnel propagator~\cite{BornWolf}.
\detectorGenerator completes the cycle and works as an auxiliary generator, mapping the propagated intensity \propagatedIntensity to the measured detector intensity $\detectorIntensity=
\detectorGenerator\propagatedIntensity=
\detectorGenerator|\propagator\objectWaveField|^2
$ using a single channel for the input and output.
Due to the propagator \propagator, \detectorGenerator does not need to learn the well-known physical process; thus it only learns the experimental effects of the intensity measurements, \eg the point-spread function and flat-field artifacts.
Finally, the intensity discriminator \intensityDiscriminator is used to classify intensity measurements as ``real'' or ``fake''.
For more details about the PhaseGAN architecture, see the \href{link}{Supplement 1}.

Our goal is to learn two mappings simultaneously: i) detector images to complex object wavefield $\objectGenerator: \detectorIntensity \rightarrow \objectWaveField$, and ii) propagated diffraction patterns to detector images $\detectorGenerator: \propagatedIntensity \rightarrow \detectorIntensity $.
This goal is achieved by optimizing
\begin{alignat}{2}
\smash{
    \underset
    {\objectGenerator,\detectorGenerator}
    {\operatorname{arg\,min}}\
    \underset
    {\phaseDiscriminator,\intensityDiscriminator}
    {\operatorname{arg\,max}}
}\
&
&&\ganLoss
(\objectGenerator, \detectorGenerator, \phaseDiscriminator, \intensityDiscriminator) + \nonumber\\
&\cycleLossWeight
&&\cycleLoss
(\objectGenerator, \detectorGenerator) +\nonumber\\
&\frcLossWeight
&&\frcLoss
(\objectGenerator, \detectorGenerator)
.
\label{eq:loss_tot}
\end{alignat}

This objective is a combination of three terms: an adversarial term, a cycle consistency term, and a \ac{FRC} term.
The relative weight of the cycle consistency and \ac{FRC} losses with respect to the adversarial loss is parametrized by \cycleLossWeight and \frcLossWeight, respectively.
The schematic of the learning process is depicted in Fig.~\ref{fig:Main}.

The first term \ganLoss of Eq.~(\ref{eq:loss_tot}) is the adversarial loss~\cite{goodfellow2014generative}
\begin{equation}
\begin{alignedat}{1}
 \ganLoss&(\objectGenerator, \detectorGenerator, \phaseDiscriminator, \intensityDiscriminator) = \\
&\mathbb{E}_{\objectWaveField \sim \phaseSamples}
[\log(\phaseDiscriminator(\objectWaveField))]+ \\
&\mathbb{E}_{\objectWaveField \sim \phaseSamples}
[\log(1-\intensityDiscriminator(\detectorGenerator|\propagator\objectWaveField|^2))] +\\
&\mathbb{E}_{\detectorIntensity \sim \intensitySamples}
[\log(\intensityDiscriminator(\mathbf \detectorIntensity))] + \\
&\mathbb{E}_{\detectorIntensity \sim \intensitySamples}
[\log(1-\phaseDiscriminator(\objectGenerator(\mathbf \detectorIntensity)))]
.
\label{eq:loss_VX}
\end{alignedat}
\end{equation}
In Eq.~(\ref{eq:loss_VX}), $\mathbb E_{\mathbf x\sim\mathcal X}$ denotes the expectation of the distribution $\mathcal X$, and \phaseSamples and \intensitySamples are the phase and intensity distributions, respectively.

The second term (\cycleLoss) of Eq.~(\ref{eq:loss_tot}) requires cycle consistency to confine generator outputs so that it is not just creating random permutation of images following the same data distribution from the desired dataset.
As shown in Fig.~\ref{fig:Main}, regardless of where we start the loop we should end up at the starting point, \ie $\objectGenerator(\detectorGenerator|\propagator\objectWaveField|^2) = \objectWaveField$ and $\detectorGenerator|\propagator(\objectGenerator(\detectorIntensity))|^2 = \detectorIntensity$. This cycle consistency loss can be expressed as:
\begin{equation}
\begin{alignedat}{2}
\cycleLoss(\objectGenerator, \detectorGenerator) =
 &\mathbb{E}_{\objectWaveField \sim \phaseSamples}
 &&[\| \objectGenerator(\detectorGenerator|\propagator\objectWaveField|^2) - \objectWaveField  \|_1] +\\
 &\mathbb{E}_{\detectorIntensity \sim \intensitySamples}
 &&[ \| \detectorGenerator|\propagator(\objectGenerator(\detectorIntensity))|^2 - \detectorIntensity  \|_1]
 .
\label{eq:loss_cyc}
\end{alignedat}
\end{equation}
The last term in Eq.~(\ref{eq:loss_tot}), \frcLoss, calculates the \ac{FRC}. \ac{FRC} takes two images or complex waves and measures the normalised cross-correlation in Fourier space over rings~\cite{saxton1982correlation,VANHEEL2005250}.
Fourier ring correlation can help to avoid common frequency artifacts such as the twin-image problem~\cite{Gabor1948,Latychevskaia2007TwinImage} or missing frequencies due to the physical propagation. The \frcLoss is defined as follows:
\begin{equation}
\begin{alignedat}{2}
\frcLoss(\objectGenerator,\detectorGenerator)
&=&\mathbb{E}_{\objectWaveField \sim \phaseSamples}
&[\| 1 - \frc( \objectGenerator(\detectorGenerator|\propagator\objectWaveField|^2),\objectWaveField) \|_2] \\
&+&\mathbb{E}_{\detectorIntensity \sim \intensitySamples}
&[\| 1 - \frc(\detectorGenerator|\propagator(\objectGenerator(\detectorIntensity))|^2,\detectorIntensity)  \|_2],
\label{eq:loss_frc}
\end{alignedat}
\end{equation}
where $\frc$ is the Fourier ring correlation operator that calculates the \ac{FRC} over all the Fourier space rings.

\section{Validation Results}\label{sec:Validation Results}
In this section, we perform phase-retrieval experiments to validate \method{PhaseGAN}. Furthermore, we compare its performance to other state-of-the-art \ac{DL} methods. This comparison is made with synthetic data in the near-field regime.

To validate \method{PhaseGAN} and compare its performance to other \ac{DL} methods, we generate synthetic X-ray imaging experiments in the near-field regime. The synthetic training dataset consists of 10,000 complex objects and 10,000 synthetic detector images.
These sets are unpaired. However, paired solutions for the detector and object simulations are available for validation purposes and training state-of-the-art paired approaches.
The wavelength of these experiments is $\lambda=1~\textup{\AA}$, and the pixel size in the object space is constrained to 1~$\mu$m.
Objects are composed of a random number between one and $N$ of rectangles and circles over a $256\times256$ frame.
The complex wavefront of such objects is given by their transmissivity. The transmissivity is estimated by their complex index of refraction $n=1-\delta+j\beta$ and a random thickness ($t$), up to a maximum thickness ($t_{\rm{max}}$) of 10~nm.
For our simulations $\delta$ and $\beta$ are fixed to $10^{-3}$ and $10^{-6}$, respectively.
The complex wavefront after the object in the projection approximation is given by:
\begin{equation}
    \objectWaveField({\bm r})={\bm \psi_i} \exp\big(j k n t({\bm r})\big)~,
    \label{eq:transmissivity}
\end{equation}
where ${\bm \psi_i}$ is the illumination wavefront at the object plane, $k=2\pi/\lambda$ is the wavenumber, ${\bm r}$ are the frame coordinates, and $t({\bm r})$ is the frame thickness map.
Then, this wavefront is propagated to the detector ($\propagator\objectWaveField$) using the near-field propagator.
The near-field detector has an effective pixel-size equal to 1~$\mu$m (equal to the sample-simulated pixel size) and is assumed to be 10~cm away from the sample.
We also include flat-field noise, \ie variable ${\bm \psi_i}$ for each frame. This flat-field noise is simulated with 15 elements of a basis extracted by \ac{PCA} from MHz-imaging data coming from the European \ac{XFEL}~\cite{Vagovic2019MHz}. Examples of the simulated holograms can be found in the \href{link}{Supplement 1}.
We assume that the detector has photon counting capabilities; thus, the noise has Poissonian behaviour. The amount of photons simulated per frame is approximately $6.6\cdot 10^7$ photons.

We compare the performance of \method{PhaseGAN} to three other methods.
The first is a classic supervised learning approach using \method{paired} datasets and an $L_2$ loss, as used by most current phase-retrieval approaches.
The second uses the same architecture as before, but with additional adversarial terms as in \method{pix2pix}~\cite{isola2017image}. The global loss function in this \method{pix2pix} method is defined by:
\begin{eqnarray}
\mathcal{L}(\objectGenerator,\phaseDiscriminator) &=& \mathbb{E}_{\objectWaveField \sim \phaseSamples}
[\log(\phaseDiscriminator(\objectWaveField))]\label{eq:pix2pixloss}\\&+&
\mathbb{E}_{\detectorIntensity \sim \intensitySamples}
[\log(1-\phaseDiscriminator(\objectGenerator(\mathbf \detectorIntensity)))]\nonumber\\
&+& \mseLossWeight \mathbb{E}_{(\objectWaveField,\detectorIntensity)\sim (\phaseSamples,\intensitySamples)}\| \objectGenerator(\detectorIntensity) -\objectWaveField\|_2~. \nonumber
\end{eqnarray}
The first two terms of Eq.~(\ref{eq:pix2pixloss}) calculate the adversarial loss in a similar way as we defined \ganLoss in Eq.~(\ref{eq:loss_VX}).
The weight of the $L_2$ loss, \mseLossWeight, was set to 100.
The third method is the standard \method{CycleGAN} approach presented in Fig.~\ref{fig:Schema}(b).
We use the same global loss function as expressed in Eq.~(\ref{eq:loss_tot}), but without including the physics of the image formation (\propagator) as in Eqs.~(\ref{eq:loss_VX}), (\ref{eq:loss_cyc}), and (\ref{eq:loss_frc}).
For the training of \method{CycleGAN}, we found that the optimal performance was obtained when $\cycleLossWeight = 20$ and $\frcLossWeight = 4$, with an additional weight of 2.5 on the first terms of Eqs. (\ref{eq:loss_cyc}) and (\ref{eq:loss_frc}).
For the \method{PhaseGAN} training, we set $\cycleLossWeight = 20$ and $\frcLossWeight = 10$.
For all experiments, we use the same phase-retrieval network $\objectGenerator$ and the same training dataset.
The dataset was paired for the training of the first two methods, but unpaired for the training of  \method{CycleGAN} and \method{PhaseGAN}.
The ADAM optimizer~\cite{kingma2014adam} with a batch size of 16 was used throughout the training.
The generator learning rates were set to be 0.0002 for all four methods.
For \method{pix2pix}, \method{CycleGAN}, and \method{PhaseGAN}, the discriminator learning rates were set to be 0.0001.
We decayed all learning rates by 10 every 30 epochs and stopped training after 70 epochs.

The phase-retrieved results are quantified by using $L_2$ norm, \ac{DSSIM}~\cite{Wang2004SSIM}, and \ac{FRCM}.
\ac{FRCM} calculates the mean square of the difference between the Fourier ring correlation and unity over all spatial frequencies.
Thus, smaller \ac{FRCM} values imply a higher similarity between two images.
Please note that such metrics are only partially able to capture the ability of a \ac{GAN} to produce data distribution samples \cite{blau2018perception}.
It must also be considered that while these metrics assume the reference solution to be available, it is ---for our method and \method{CycleGAN}--- only used to compute the metric, never in training.
For qualitative assessment, a reader is referred to Tbl.~\ref{tbl:Results}.
Tbl.~\ref{tbl:Results} depicts the real and imaginary part of a zoom-in area of one of the validation samples or oracle and the retrieved results for each method.
In Tbl.~\ref{tbl:Results}, we also report, for each of the four \ac{DL} methods, the logarithmic frequency distribution and the average value ($\mu$) for the aforementioned validation metrics over 1000 validation images.
More information about the statistical distribution of the metric values and line profiles through different validation images can be found in the \href{link}{Supplement 1}.

\begin{table}[htbp!]
\centering
\caption{Comparison of different methods \emph{(rows)} applied to the same input according to different metrics \emph{(columns)}.
\label{tbl:Results}}
\setlength{\tabcolsep}{4pt}
\begin{tabular}{crrrrrrr}
\hline
  &
  \multicolumn2c{Example patch}&
  \multicolumn1c{$L_2$}&
  \multicolumn1c{DSSIM}&
  \multicolumn1c{FRCM}\\
  & \multicolumn1c{\footnotesize Real}
  & \multicolumn1c{\footnotesize Imaginary}
  &
  {\footnotesize $\times 10^{-5}$}&
  {\footnotesize $\times 10^{-4}$}&
  {\footnotesize $\times 10^{-4}$}\\
\hline
\methodInTable{oracle}
 \multicolumn1c{---}&
 \multicolumn1c{---}&
 \multicolumn1c{---}\\
\methodInTable{paired}
\minihisto{4.20}{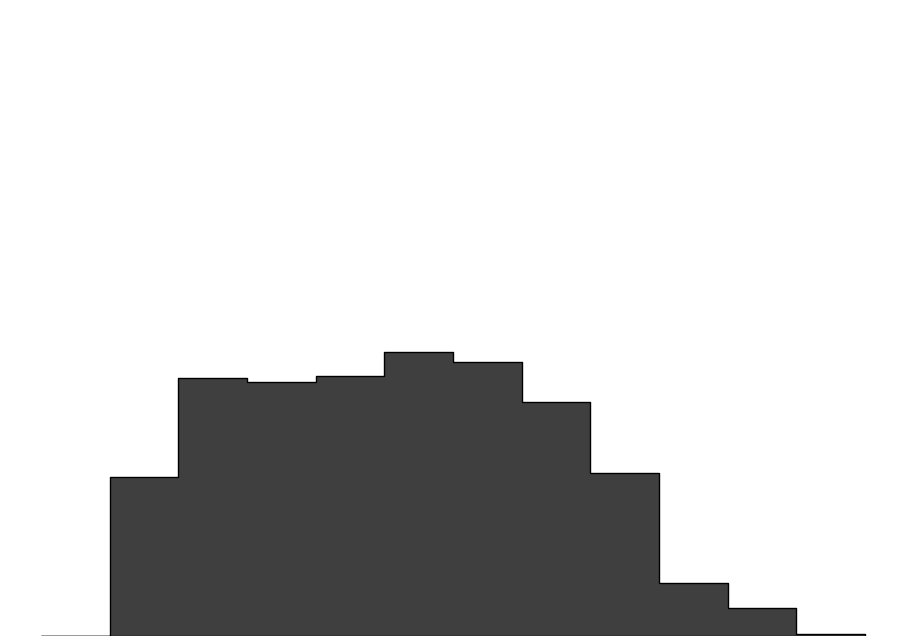}&
\minihisto{4.59}{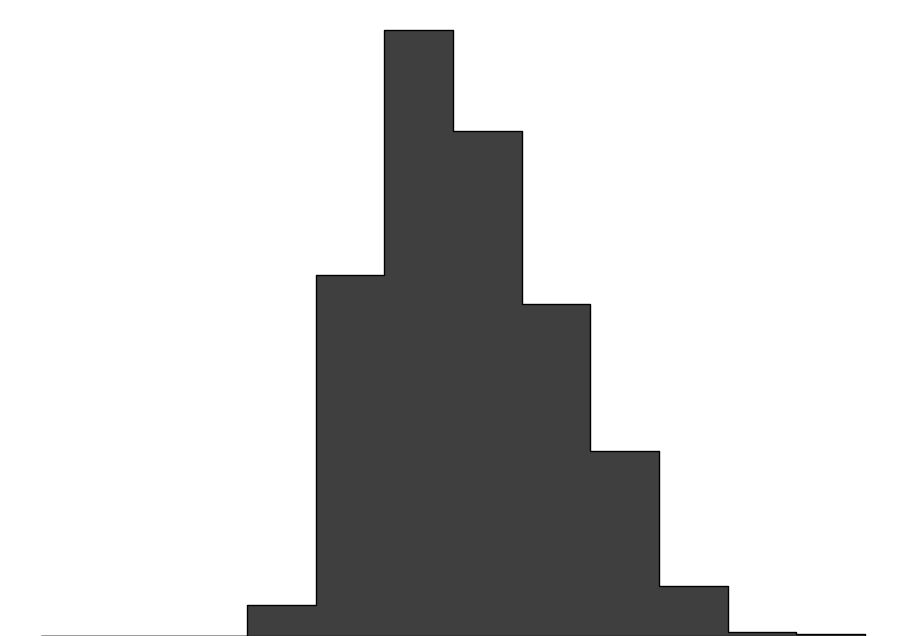}&
\minihisto{2.56}{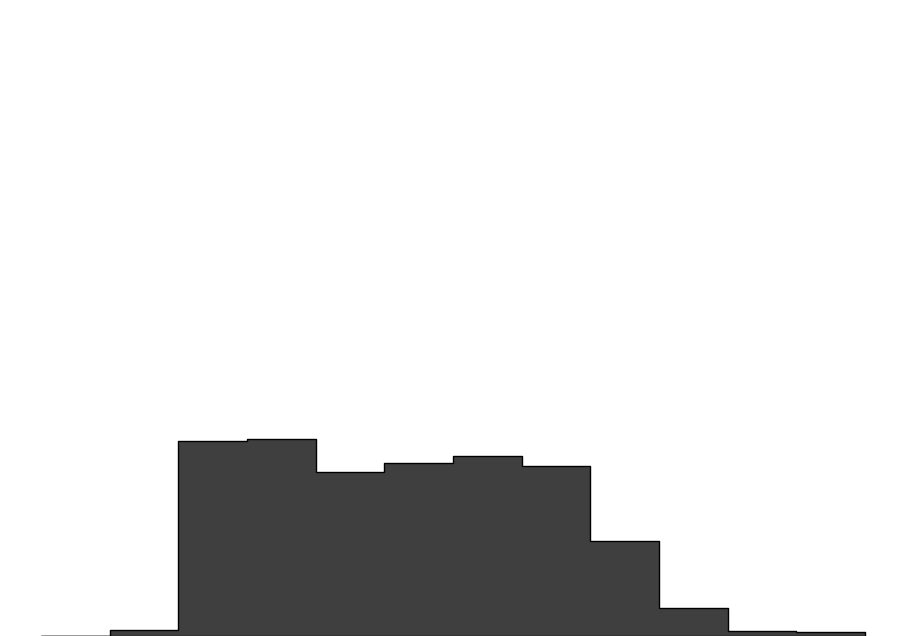}
\\
\methodInTable{pix2pix}
\minihisto{4.03}{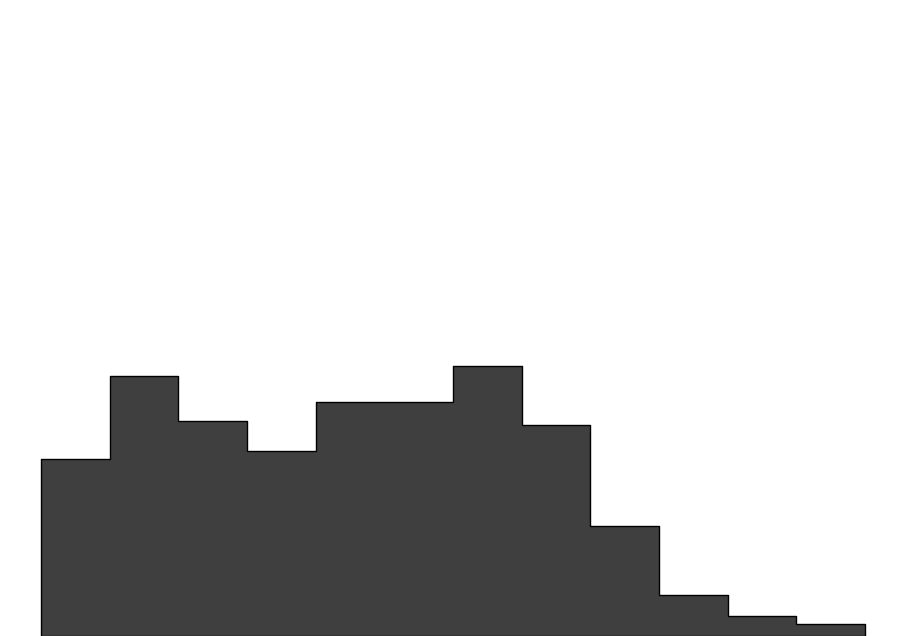} &
\minihisto{3.09}{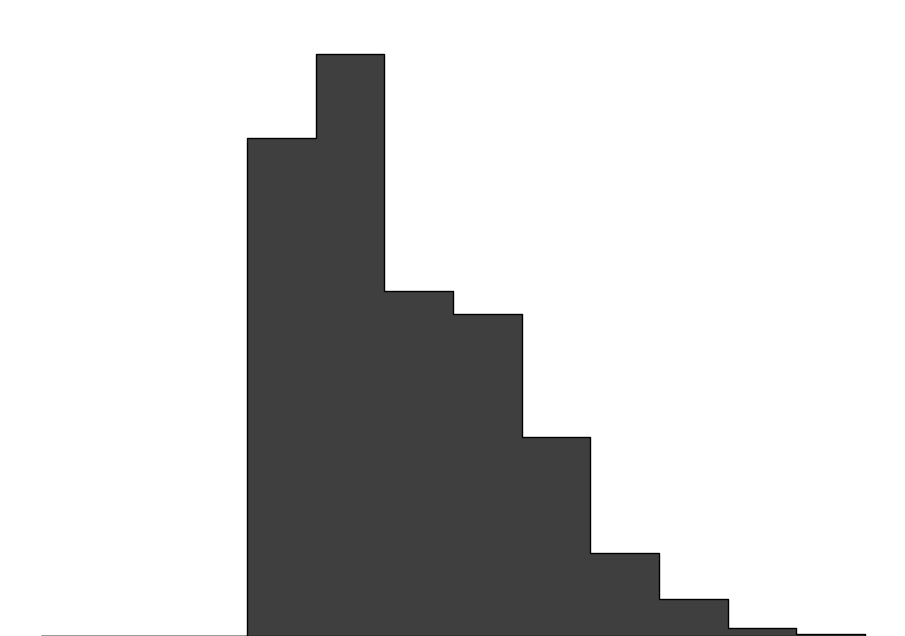} &
\minihisto{1.08}{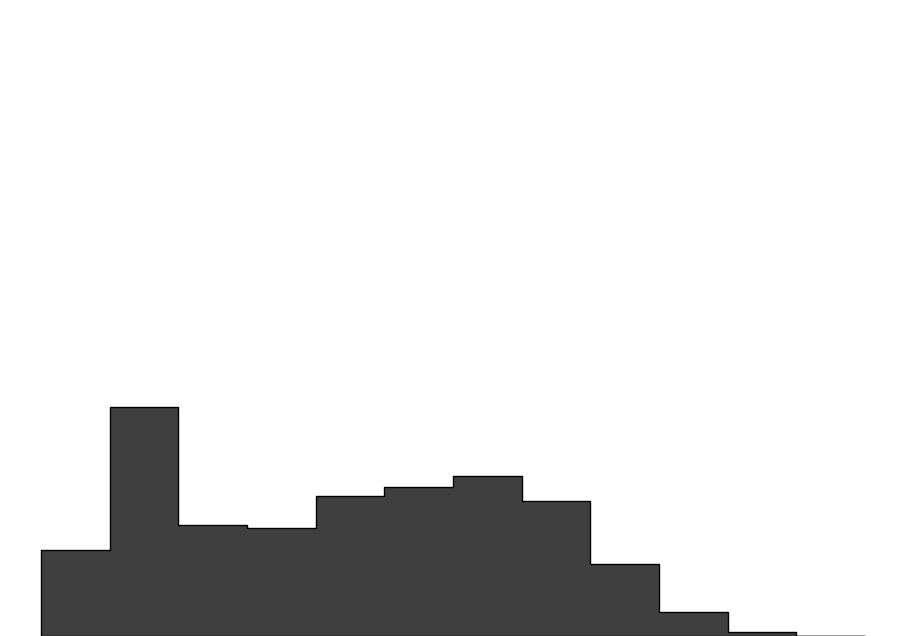}
\\
\methodInTable{CycleGAN}
\minihisto{28.0}{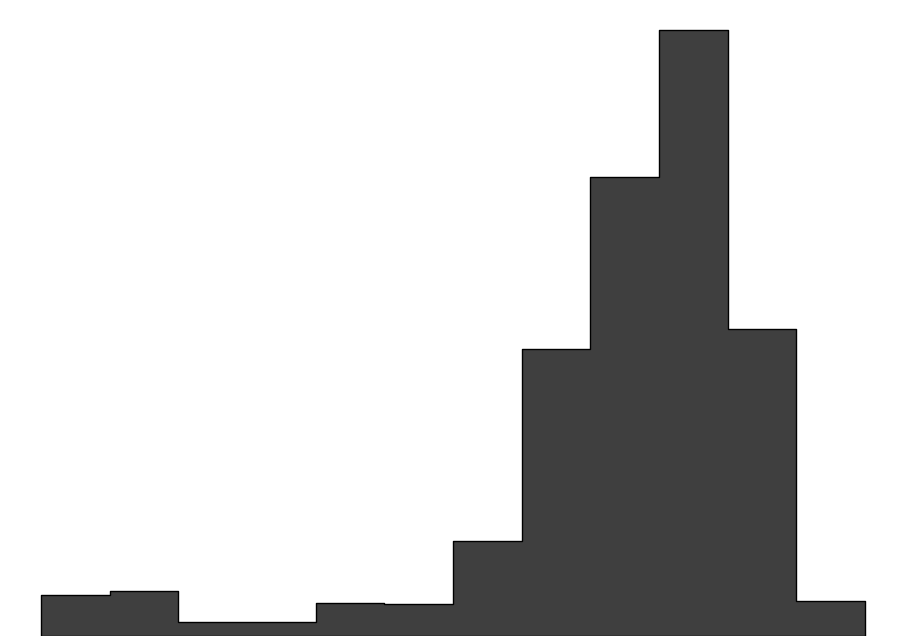}&
\minihisto{27.5}{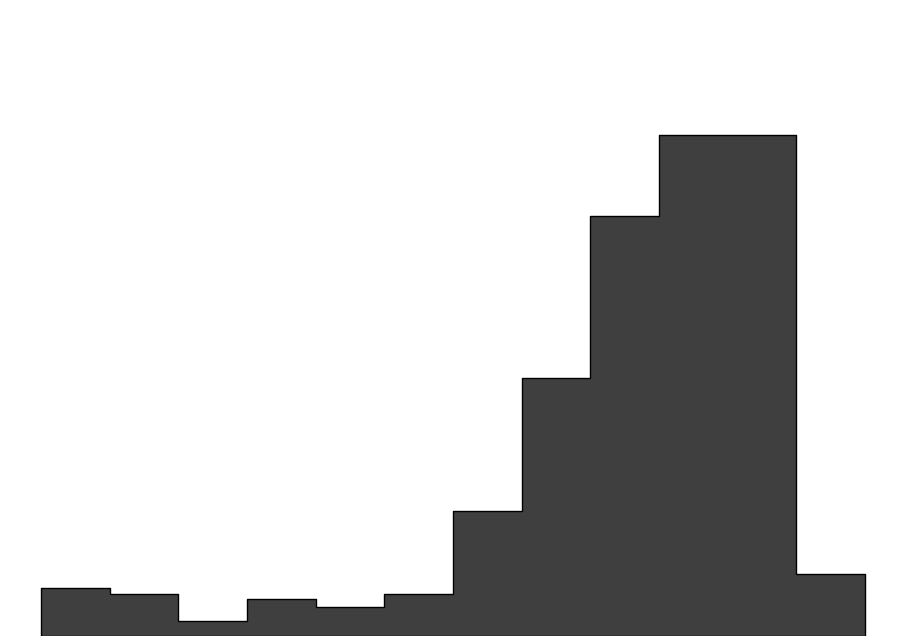}&
\minihisto{94.8}{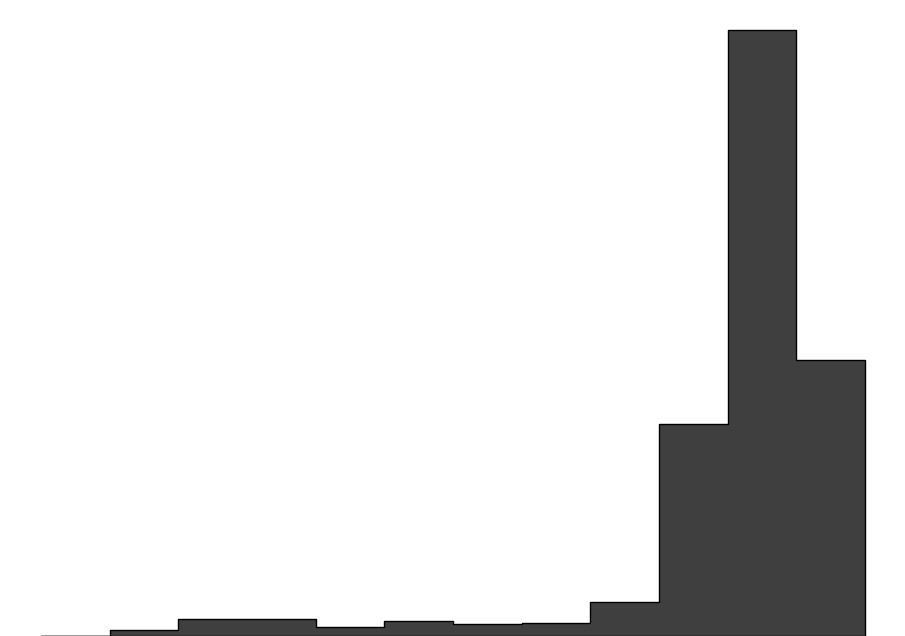}
\\
\methodInTable{PhaseGAN}
\minihisto{7.37}{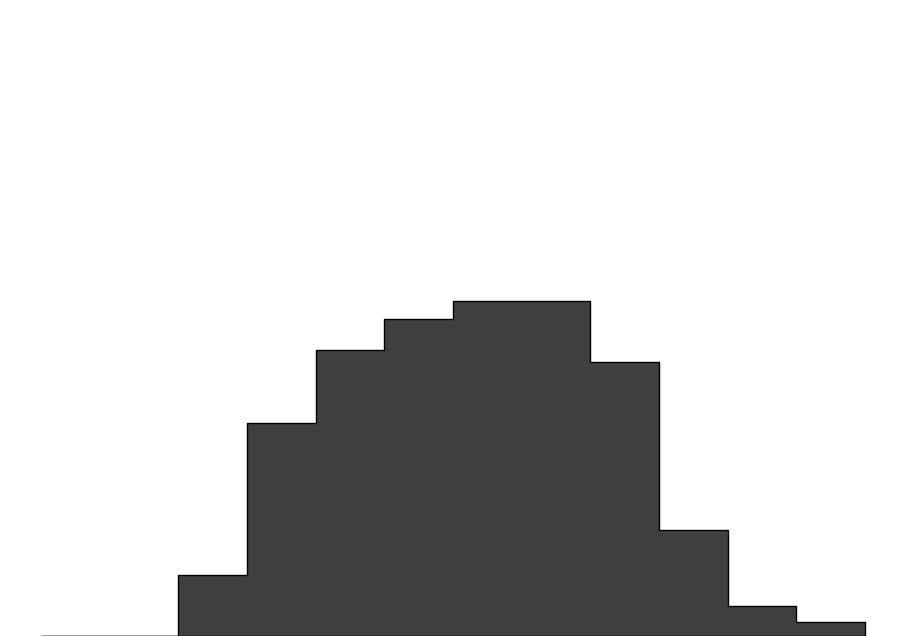}&
\minihisto{8.91}{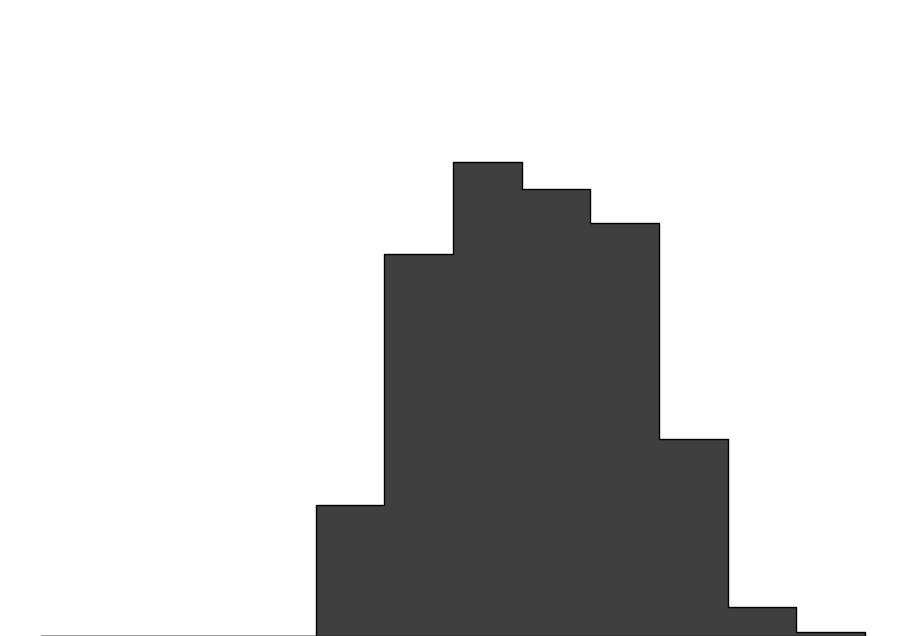}&
\minihisto{2.54}{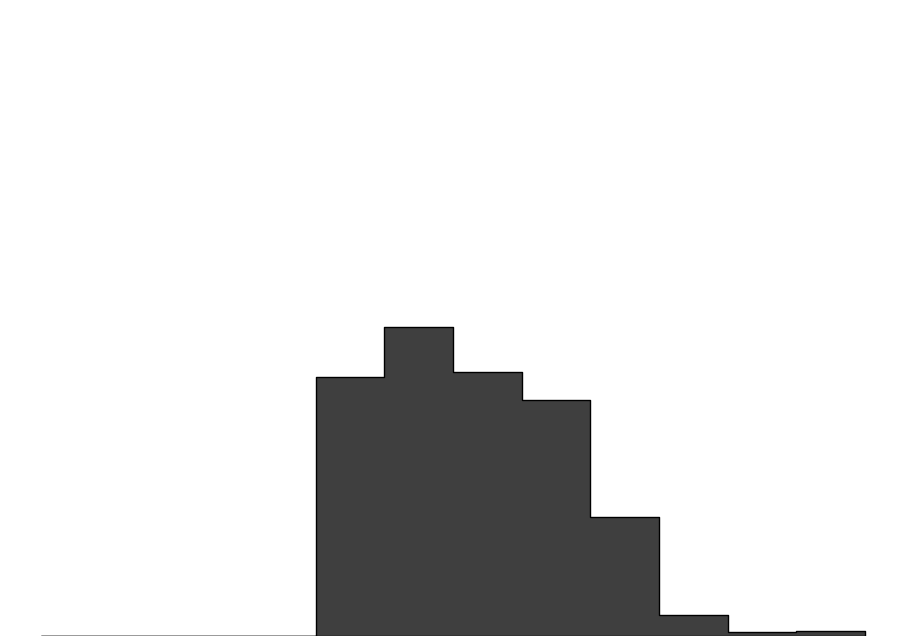}
\\
\hline
\end{tabular}
\end{table}

\begin{figure*}[htb]
\centering\includegraphics*[width = \linewidth]{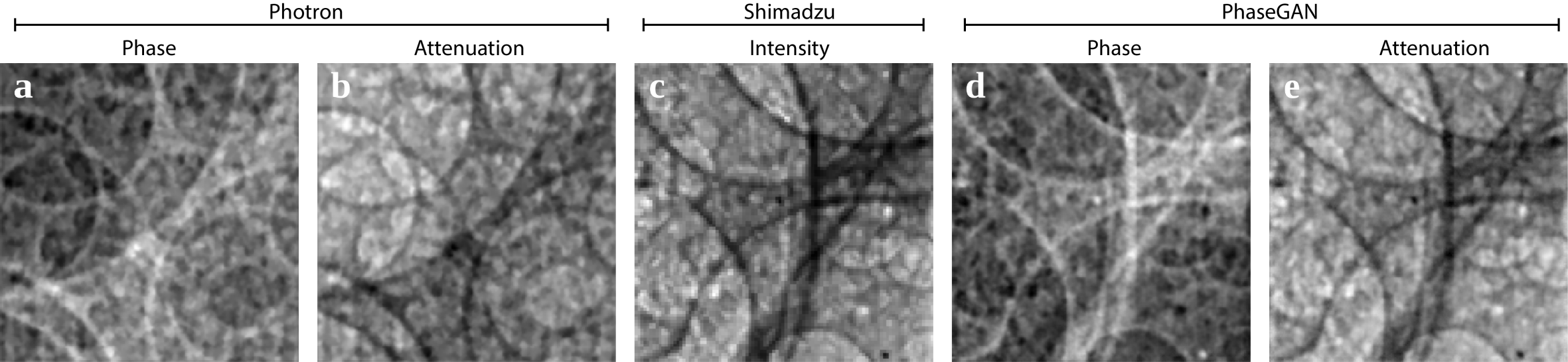}
\vspace{-.2cm}
\caption{Application to metallic foam. Phase (a) and attenuation (b) reconstructed using TIE of a frame acquired with the Photron-based system. (c) Intensity measured with the Shimadzu recording system using a single pulse. (d) Phase reconstructed and (e) attenuation retrieved by \method{PhaseGAN}.}
\label{fig:Foam}
\end{figure*}

\section{Experimental Results}
In this section, we applied \method{PhaseGAN} to experimental data recorded at the \ac{APS}, where unpaired data of metallic foams was recorded with two different detectors at independent sensing experiments.

\method{PhaseGAN} offers the opportunity to obtain phase information when phase reconstructions are not possible. To demonstrate this, we performed time-resolved X-ray imaging experiments of the cell-wall rupture of metallic foams at the Advanced Photon Source (APS)~\cite{Fezzaa2008Ultrafast}.
The coalescence of two bubbles caused by the cell-wall rupture is a crucial process, which determines the final structure of a metallic foam~\cite{Garcia-Moreno2012Coalescence}.
This process can happen within microseconds; thus, MHz microscopic techniques are required to explore it.
For this reason, we performed ultra-fast experiments with an X-ray imaging system based on a Photron FastcamSA-Z with 2~$\mu$m effective pixel size.
The Photron system acquires the cell-wall rupture movies at a frame rate of 210~kHz, which integrated over 31 pulses of APS.
Although the images acquired by the Photron camera used a few pulses, they had good contrast, which allows obtaining meaningful phase reconstructions.
Images acquired by the Photron system were interpolated to an effective pixel size of 1.6~$\mu$m and filtered using 100 iterations of a total variation denoising algorithm~\cite{TVL1-MATLAB} with denoising parameter $\lambda=1.5$.
Images obtained were phase-reconstructed using a \ac{TIE} approach for single-phase materials~\cite{Paganin2002a} assuming X-ray photons of 25.7~keV, $\delta/\beta=10^3$ and propagation distance $z=5$~mm.
A phase and attenuation reconstructions for a frame of the Photron system are shown in Fig.~\ref{fig:Foam}(a) and (b), respectively.
In order to increase the temporal resolution and to be able to use single pulses of APS, we used an X-ray MHz acquisition system based on a Shimadzu HPV-X2 camera with an effective pixel size of 3.2~$\mu$m.
This system was used to record movies of dynamic phenomena in liquid metallic foams using single pulses provided by APS with a repetition frequency of 6.5~MHz.
An example of a frame recorded with this system is shown in Fig.~\ref{fig:Foam}(c).
However, the contrast and noise were not sufficient to perform phase reconstructions with current approaches.

To overcome the impossibility of performing phase reconstructions using the frames recorded by the Shimadzu system, we used \method{PhaseGAN}.
The dataset for \method{PhaseGAN} training consists of 10000 Photron frames and 10000 Shimadzu frames, with frame sizes of $480 \times 200$ and $128 \times 128$ pixels, respectively.
Due to the different pixel sizes in the two imaging systems, the two sets of images were cropped to $200 \times 200$ and $100 \times 100$ before feeding them into the \ac{NN}.
This was done to match the field-of-view in the two different imaging domains.
We performed data augmentation by applying random rotations and flips to the randomly cropped training images to take full advantage of \method{PhaseGAN}'s capabilities.
As is commonly used in supervised learning, data augmentation is also indispensable in unsupervised approaches for the neural network to learn the desired robustness properties \cite{dosovitskiy2014discriminative}, especially when only limited training examples are available.
In our case, the holograms were captured by kHz to MHz camera systems, making detector frames very similar to each other.
\method{PhaseGAN} reconstructions without data augmentation will not learn the desired mappings from one domain to the other but only remember the common features in each frame.
The cropped Photron and Shimadzu frames were subsequently padded during the training to $256 \times 256$ and $128 \times 128$, respectively.
We slightly modified the network architecture of \method{PhaseGAN} for the training of metallic foams, where an extra step of transposed convolution was added to the expanding step in \objectGenerator to double the size of the output images due to the half-pixel size of the Photron detector in respect to the Shimadzu one.
Conversely, the last transposed convolutional layer of the \detectorGenerator was replaced by a normal convolutional layer to accommodate the double-pixel size of the Shimadzu detector with respect to the Photron detector.
We set $\cycleLossWeight = 150$ and $\frcLossWeight = 10$.
The ADAM optimizer with the same learning rates used for the synthetic data and a batch size of 40 was adopted for the metallic foam training.
The training was stopped after 100 epochs.
The \method{PhaseGAN} phase and attenuation outputs for the Shimadzu frame depicted in Fig.~\ref{fig:Foam}(c) are shown in Fig.~\ref{fig:Foam}(d) and (e), respectively.
A complete movie of the cell-wall rupture of a metallic foam (FORMGRIP alloy~\cite{Gergely2000Formgrip}) and its phase and attenuation reconstruction using \method{PhaseGAN} are provided in the supplemental  \href{https://osapublishing.figshare.com/s/95f9cfbaea9a9314c311}{Visualization 1},
\href{https://osapublishing.figshare.com/s/20a66707b78822586f0c}{2}, and
\href{https://osapublishing.figshare.com/s/98aa02ab61cfbdf2097b}{3}.
It is noticeable from the movie clip that the coalescence of the two bubbles was finished within 10 $\mu$s.
In total, 24.4 ms were consumed to reconstruct the 61 frames of the movie, \ie \method{PhaseGAN} reconstructions took 0.4~ms per frame.
Thus, PhaseGAN offers an opportunity for real-time analysis.

\section{Discussion}
\label{sec:discussion}

We have presented PhaseGAN, a novel \ac{DL} phase-retrieval approach. PhaseGAN, when compared to other approaches, provides for the first time phase reconstructions of unpaired datasets. The cyclic structure of PhaseGAN allows to include the physics of image formation in the learning loop, which further enhances the capabilities of unpaired \ac{DL} approaches, such as CycleGAN.
Although we did not include typical constraints used in iterative phase-retrieval approaches, such as support, histogram constraints, and sample symmetries, PhaseGAN performs at the level of state-of-the-art \ac{DL} phase-reconstruction approaches.
However, PhaseGAN's cyclic approach could be adapted to include such constraints to enhance its capabilities further.
Another key ingredient of PhaseGAN is the inclusion of a \ac{FRC} loss term, which penalizes common phase-reconstruction artifacts easy to filter in the Fourier domain, such as missing frequencies and the twin-imaging problem~\cite{Gabor1948,Latychevskaia2007TwinImage}.

We have demonstrated PhaseGAN's capabilities by performing near-field holographic experiments and compared the results to i) state-of-the-art paired approaches, ii) a \ac{GAN} method following the pix2pix approach, and iii) CycleGAN.
The results of the experiments, using the same training datasets, paired when needed, and phase-retrieval generator ($\objectGenerator$), demonstrate the unique capabilities of PhaseGAN. These results are reported in Table~\ref{tbl:Results}.
From this table, we can conclude that both paired approaches retrieve competitive phase reconstructions quantitatively and qualitatively. CycleGAN, due to the challenge of training on unpaired datasets, clearly performs worse than paired approaches.
PhaseGAN, although unpaired as well, retrieves results at the level of paired-training approaches.

We have applied PhaseGAN to time-resolved X-ray imaging experiments using single pulses of a storage ring to study the cell-wall rupture of metallic foams.
In this imaging modality, noisy images with low contrast and low resolution are recorded due to the limited number of photons per pulse.
This acquisition scheme records images that cannot be phase-reconstructed.
However, such an approach opens the possibility to record dynamics at MHz frame rates.
In parallel, we acquired a less noisy and better-contrast dataset that allowed phase reconstructions.
This dataset was obtained by integrating over 31 pulses and had about half of the pixel size of the time-resolved dataset.
By training using these two different sensing experiments on different realizations of metallic foam, we demonstrate the capability of PhaseGAN to produce phase reconstructions, which are not possible using any current approach.

\section{Conclusions}
\label{sec:conclusions}
To conclude, we have presented a novel cyclic \ac{DL} approach for phase reconstruction, called PhaseGAN.
This approach includes the physics of image formation and can use unpaired training datasets to enhance the capabilities of current \ac{DL}-based phase-retrieval approaches.
We have demonstrated the unique capabilities of PhaseGAN to address the phase problem when no phase reconstructions are available, but good simulations of the object or data from other experiments are.
This will enable phase reconstructions that are not possible today by correlating two independent experiments on similar samples.
For example, it will open the possibility of phase reconstructions and denoising with X-ray imaging from low-dose in-vivo measurements by correlating them with higher-dose and lower-noise measurements performed on ex-vivo samples of similar tissues and structures.
It has the potential to denoise and reconstruct the phase of time-resolved experiments to track faster phenomena with a limited number of photons per frame.

The PhaseGAN code is available at \href{https://github.com/pvilla/PhaseGAN.git}{GitHub}.

\section{Funding}
Bundesministerium für Bildung und Forschung (BMBF) (05K18KTA); Vetenskapsrådet (VR)
(2017-06719).

\section{Acknowledgments}
We are greatful to Z. Matej for his support and access to the GPU-computing cluster at MAX~IV. The presented research used resources of the Advanced Photon Source, a U.S. Department of Energy (DOE) Office of Science User Facility operated for the DOE Office of Science by Argonne National Laboratory under Contract No. DE-AC02-06CH11357. We also gratefully acknowledge the support of NVIDIA Corporation with the donation of a Quadro P4000 GPU used for this research.

\section{Disclosures}
The authors declare no conflicts of interest.

\bigskip \noindent See \href{link}{Supplement 1} for supporting content.

\bibliography{paper}

\newpage
\appendix
\renewcommand{\thesection}{S\arabic{section}}
\onecolumn

\end{document}


\maketitle

\section{PhaseGAN Architecture}

 \begin{figure}[htb]
\centering\includegraphics*[width = \linewidth]{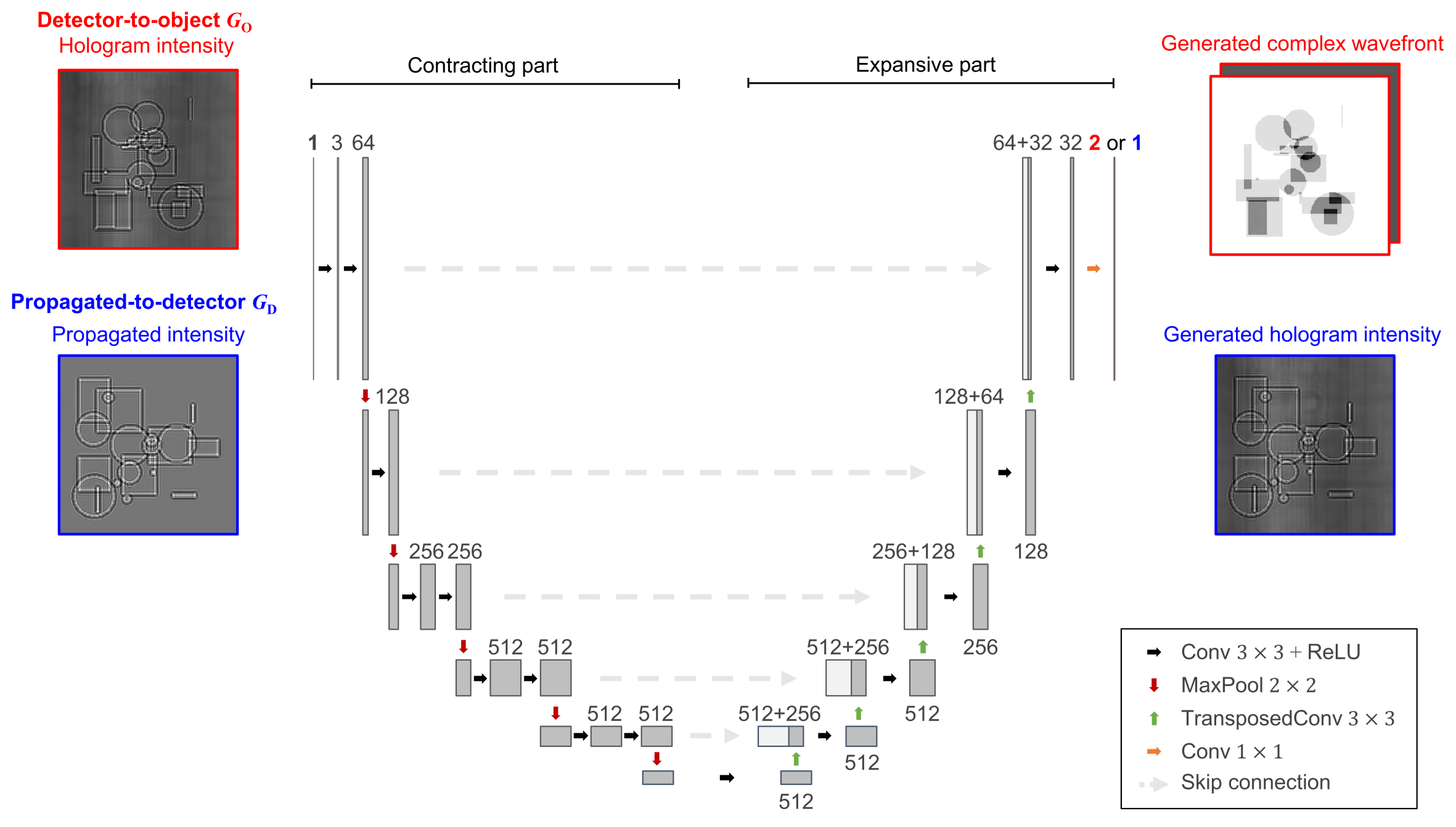}
\vspace{-.2cm}
\caption{Generator architecture of PhaseGAN. The contracting and expansive path of the two PhaseGAN generators (\objectGenerator and \detectorGenerator) are depicted here. The phase-retrieval generator \objectGenerator maps the hologram intensity in the detector domain to a complex wavefront in the object domain, generating a two-channel output from a single channel input. \detectorGenerator maps the propagated intensity in the detector plane to the measured detector intensity, \ie it maps one input channel to one output channel.}
\label{fig:PhaseGAN_network}
\end{figure}

This section describes the architecture used for PhaseGAN.

The generators used in PhaseGAN are U-Net~\cite{ronneberger2015u} type end-to-end fully convolutional neural networks.
As shown in Fig.~\ref{fig:PhaseGAN_network}, the generator architecture consists of a contracting and expansive path.
In the contracting path, the spatial resolution is reduced, and the feature information is increased.
The contracting path in our model contains multiple convolutional layers with kernel size 3 $\times$ 3, each followed by a ReLU activation function.
Max pooling operations with kernel size 2 $\times$ 2 are applied to 5 of the convolutional layers. After each max pooling, the image size is reduced by 2, decreasing from 256 $\times$ 256 to 8 $\times$ 8 pixels in the lowest resolution. The number of feature layers is doubled after each pooling operation.
The extracted feature information is relocalized in the expansive path by combining upsampled feature mapping with the skip-connected high-resolution components from the contracting path.
In the expansive path, the resolution of the images is recovered by repeated application of transposed convolutions. The transposed convolution outputs are then concatenated with the associated feature map from the contracting path and then send into corresponding convolutional layers.
The generator weights are initialized by a pre-trained VGG11 encoder to improve model performance and accelerate the training process~\cite{iglovikov2018ternausnet}.

The discriminators used in this work are PatchGAN discriminators similar to the ones used in~\cite{ledig2017photo,zhu2017unpaired}. They contain four convolutional layers with 4 $\times$ 4 convolutional filters, gradually increasing the number of filters by a factor of 2 from 64 to 512. Each convolution layer is followed by a batch normalization layer and a leaky RELU activation function with a slope of 0.2. The discriminators are trained to distinguish real images from the ones faked by the generator. For an image of size 256 $\times$ 256, the discriminator output a 30 $\times$ 30 matrix, where each matrix element corresponds to a 70$\times$70 image area, examining if this part is from the training dataset or not.

The PhaseGAN architecture was trained using the MAX~IV computing cluster.
We used Nvidia Tesla V100 SXM2 GPU with 16 and 32 GB of RAM to train the synthetic and metallic foam datasets, respectively.
For a given dataset, the speed of training is dependent on various elements including the network architecture, batch size, and the memory of the devices.
For the training of metallic foam dataset using 32 GB memory and batch size of 40, it took less than 10 hours to go through 100 epochs. The reconstruction process is less time-consuming. It took 20 ms to reconstruct 50 frames.
The generators each contains 22.93 million learnable parameters, while the discriminators have 2.76 M. The model sizes of the well-trained generator and discriminator are 460 MB and 55 MB, respectively.

We provide the PyTorch implementation of PhaseGAN, which is based on the architectures from \cite{zhu2017unpaired} and \cite{iglovikov2018ternausnet}.
The PhaseGAN implementation is available at \href{https://github.com/pvilla/PhaseGAN.git}{GitHub}.

\section{PhaseGAN results summary}
This section presents the training strategy and results obtained for the validation (synthetic) and metallic foam experiments.

\begin{figure}[htb]
\centering\includegraphics*[width = \linewidth]{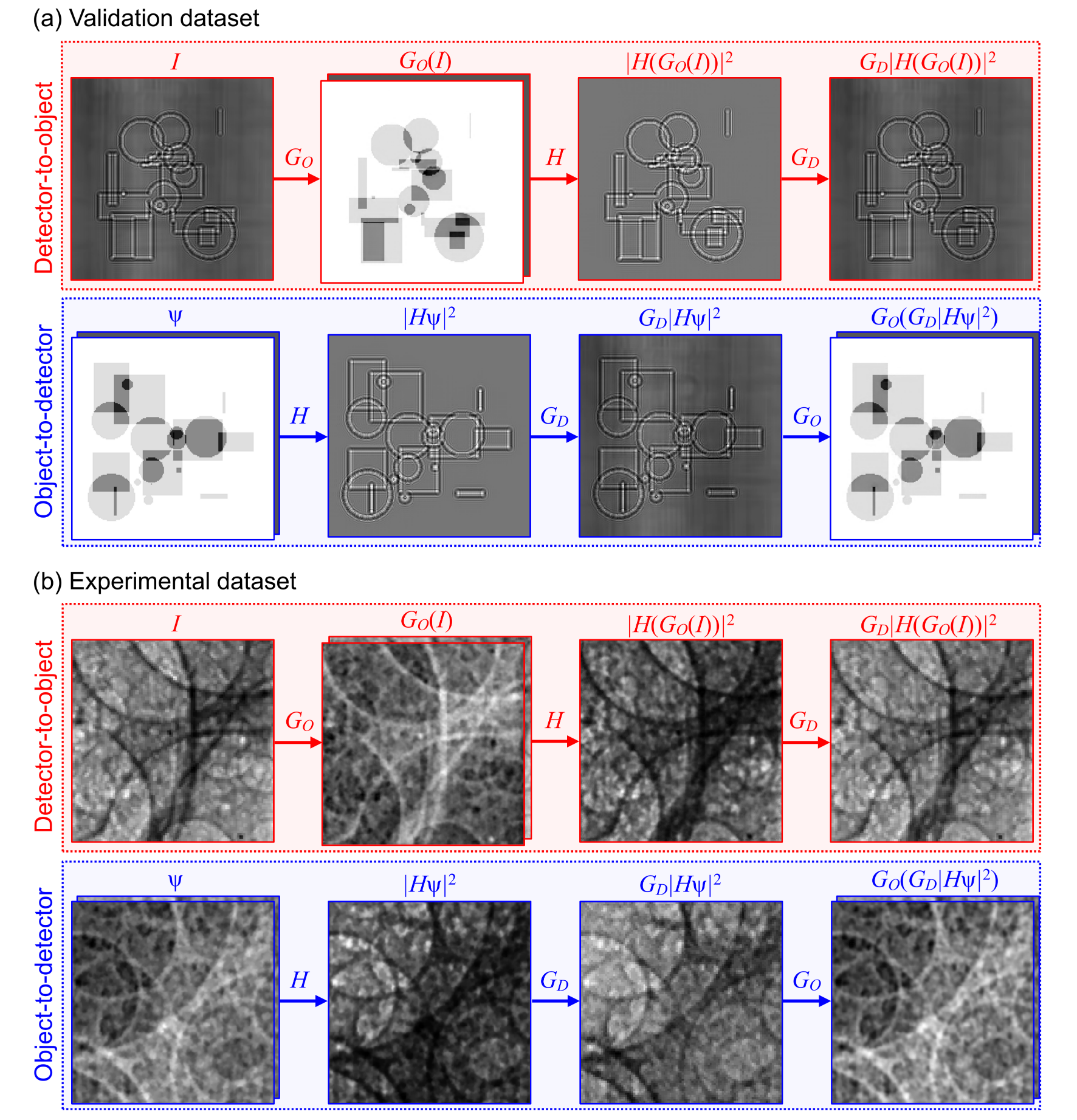}
\vspace{-.2cm}
\caption{PhaseGAN cycle-consistency illustration for the (a) validation and (b) experimental datasets. Inside the red box, the cycle from intensity measurements back to intensity measurements is shown. The blue box depicts the complex-wavefront closed cycle. The intermediate steps within each cycle are illustrated. Those intermediate steps use \objectGenerator, \propagator, and \detectorGenerator.}
\label{fig:Cycle}
\end{figure}

PhaseGAN is an unpaired phase-reconstruction approach.
To train on unpaired datasets, PhaseGAN needs two cycles that use either detector measurements or phase-reconstructed objects as input.
Each of these cycles is required to be consistent, \ie the input should be recovered at the end of the cycle.
The two PhaseGAN cycles with their intermediate steps for the validation and experimental datasets are shown in Fig.~\ref{fig:Cycle}.
These steps include two generators \objectGenerator and \detectorGenerator.
The \objectGenerator learns the mapping between the measured detector intensity ($I$) to the object complex wavefront ($\Psi$).
The \detectorGenerator learns the mapping between the estimated intensity on the detector plane to the actual measured intensity.
Another intermediate step includes the physics of the image formation via the propagator (\propagator).
\propagator propagates the complex wavefront from the object plane to the detector plane.
The inclusion of the propagator is crucial to enhance the performance of the phase-reconstructions obtained by PhaseGAN.
Finally,  Fig.~\ref{fig:Cycle} evidences the capability of PhaseGAN to fulfil the cycle consistency.

We have performed several tests to understand the capabilities of PhaseGAN compared to state-of-the-art DL approaches.
Specifically, we have compared PhaseGAN to:
i) classical supervised learning approach using \method{paired} datasets,
ii) adversarial supervised learning with paired datasets using a \method{pix2pix}~\cite{isola2017image},
and iii) standard \method{CycleGAN}~\cite{zhu2017unpaired}.
For more details about the used methods, the reader is referred to the main text.
All these approaches use the same \objectGenerator to retrieve the phase.

\begin{figure}[htb]
\centering\includegraphics*[height = 0.9\textheight]{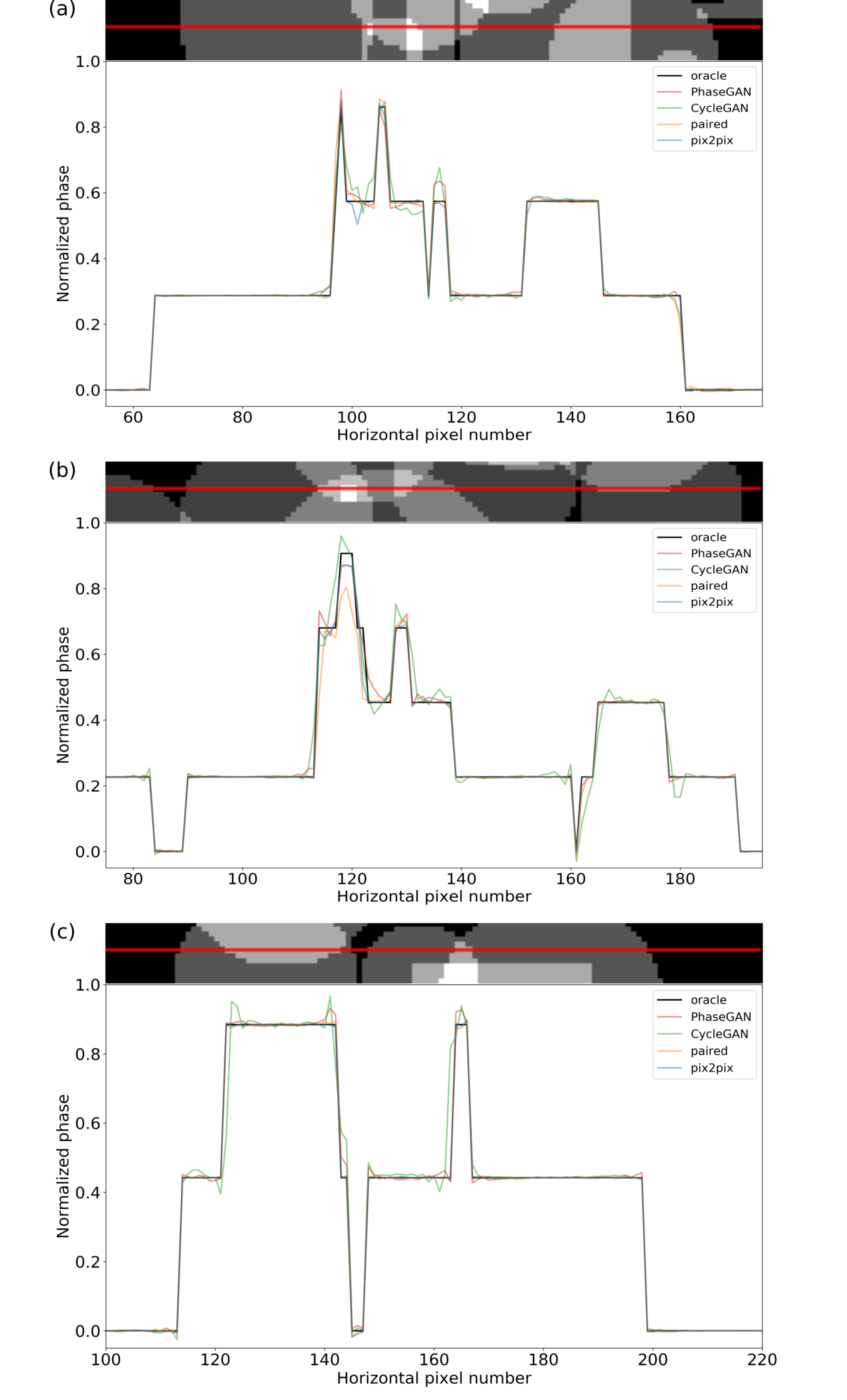}
\caption{Phase-reconstructed line profiles by the four DL methods for three independent validation samples. Segments of the oracle images are shown on top of each line-profile plots, where the red path indicates the depicted line profile.
Graph (a) shows a line profile of the validation sample patch in Tbl.~1. It crosses through the upper left overlapping area, where three circles overlap with each other. Two more examples showing details about the network reconstructions over object edges and overlapping areas are given in (b) and (c). }
\label{fig:Line_profile}
\end{figure}

One of the most simple tests to understand its capabilities was to look at phase profiles over areas difficult to reconstruct, \ie regions with a high variation of the phase profile over a small area.
The results for three line profiles are shown in  Fig.~\ref{fig:Line_profile}.
It can be seen that all four methods are capable of reconstructing the homogeneous regions seen in the reference or oracle wavefield. However, the main discrepancies were observed around the object edges.

\begin{figure}[htb]
\centering\includegraphics*[width = \linewidth]{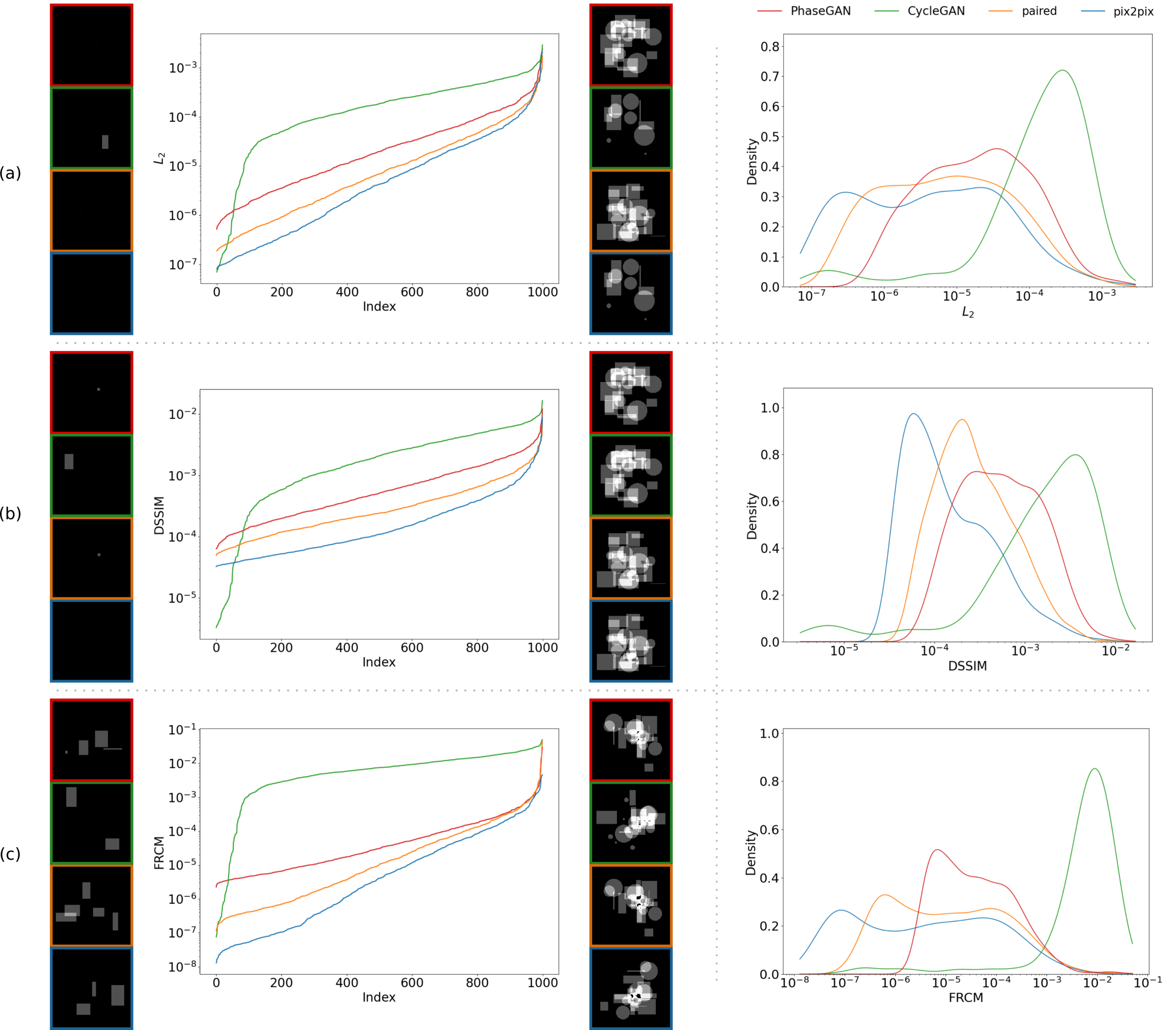}
\vspace{-.2cm}
\caption{Comparison of the ranked distribution (left) and the kernel density estimation (right) of PhaseGAN (red), CycleGAN (green), paired (orange), and pix2pix (blue) according to $L_2$ norm (a), DSSIM (b), and FRCM (c).}
\label{fig:Error}
\end{figure}

Second, we report the statistical distributions of three quality metrics $L_2$ norm, DSSIM, and FRCM for the four DL approaches.
For more details about these metrics, the reader is referred to the main text.
Smaller values of these three metrics correspond to better reconstructions.
Conversely, larger values evidence worse reconstructions.
The distributions over 1000 validation images for the $L_2$ norm, DSSIM, and FRCM, are shown in Fig.~\ref{fig:Error}(a), (b), and (c), respectively. Each validation contains a random number of objects ranging from 1 to 25. The phase of the images ranges from 0 to $\pi$ to avoid the problem of phase wrapping. For each metric, we also include the best-performed and the worst-performed validation images of each DL method.
The left side of the figure depicts the ranked distribution for each metric from smaller to larger values.
The ranked distributions are independent for each of the DL methods, \eg the smallest value for a given metric and method does not have to be obtained from the same input image as for another method with the same metric. The image patches on the left (right) side of each ranked distribution show the best (worst) phase-retrieved results for each DL approach. The frame colour follows the legend colour code for each method.
As expected, most of the methods perform better with fewer objects than with a large quantity of them. The overlap between objects also plays a role in the method's performance.
On the right side of  Fig.~\ref{fig:Error}, the kernel-density estimations are depicted for each of the methods and metrics.
These distributions are calculated over the logarithmic distribution of values to enhance the differences between the methods.
One can see that PhaseGAN outperforms CycleGAN and performs at the level of current-state-of-the-art paired DL approaches when applied to the phase problem.

\begin{figure}[htb]
\centering\includegraphics*[width = \linewidth]{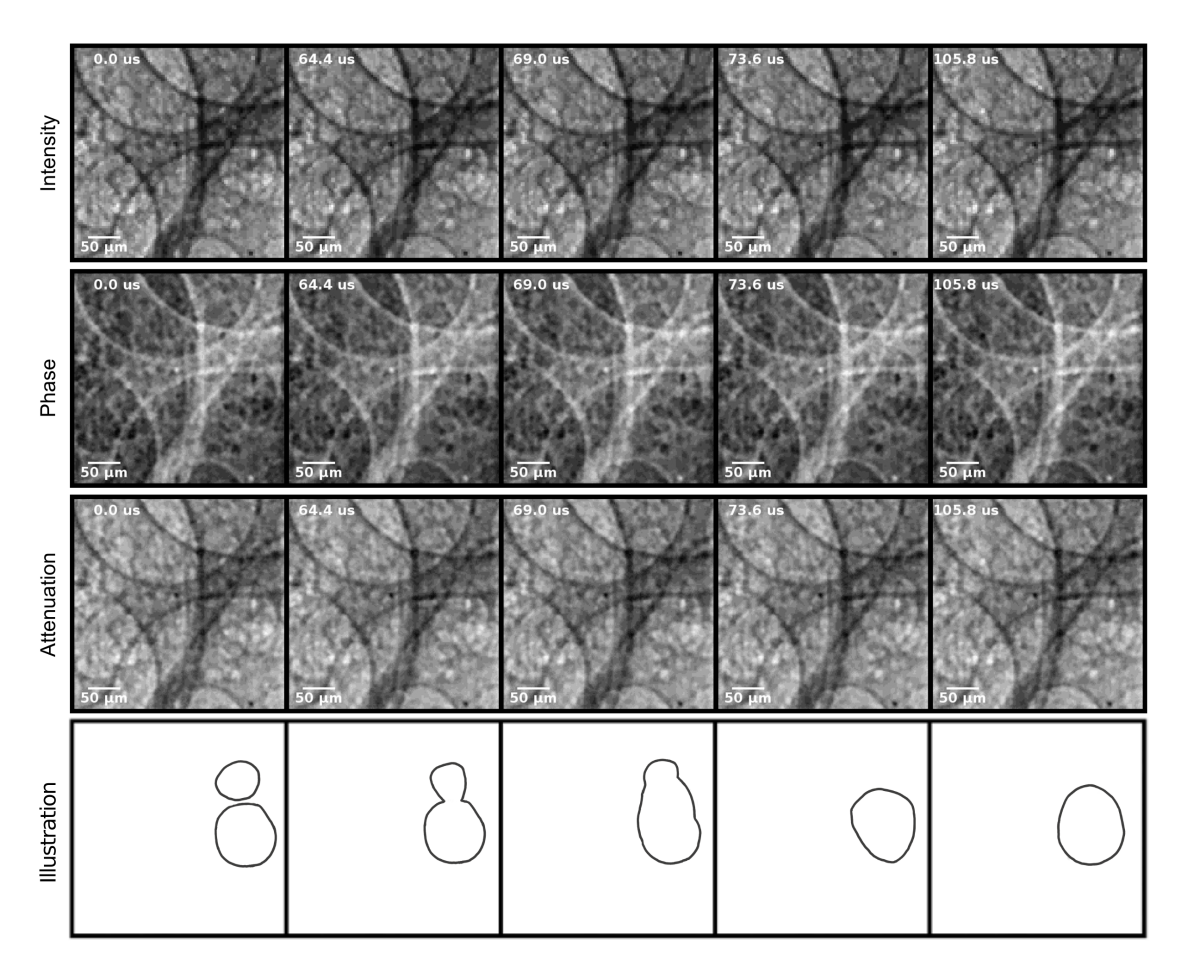}%
\vspace{-.2cm}
\caption{Clips of the supplementary movie (see \href{https://osapublishing.figshare.com/s/95f9cfbaea9a9314c311}{Visualization 1},
\href{https://osapublishing.figshare.com/s/20a66707b78822586f0c}{2}, and
\href{https://osapublishing.figshare.com/s/98aa02ab61cfbdf2097b}{3}) showing two bubble coalescence of metallic foam. The sketches on the bottom row illustrate this process.}%
\label{fig:Movie_frames}
\end{figure}

Finally, we display five selected frames extracted from a time-resolved X-ray imaging experiment in Fig.~\ref{fig:Movie_frames}.
This experiment studied the coalescence of metallic-foam bubbles.
This is a crucial process that determines the final structure of the metallic foam~\cite{Garcia-Moreno2012Coalescence}.
The Intensity row corresponds to measurements performed with a MHz X-ray imaging acquisition system based on a Shimadzu HPV-X2 camera.
This system was capable of recording single X-ray pulses provided by the Advanced Photon Source (APS).
The phase and attenuation rows correspond to the phase-retrieved results from PhaseGAN, which cannot be provided by current methods. The last row in Fig.~\ref{fig:Movie_frames} shows a schematic illustration of the coalescence process.

PhaseGAN provided a satisfactory solution for this condition, which can provide almost real-time (kHz) phase reconstructions avoiding experimental artifacts in the absence of paired image examples.
PhaseGAN can also work as an alternative to the traditional iterative phase reconstruction methods in the need for large volumes of data and rapid reconstructions.

\bibliography{supplement}